\documentclass[journal=nalefd,manuscript=letter]{achemso}
\usepackage{graphicx,subcaption,amsmath,gensymb,esvect,xcolor,xr}

\setkeys{acs}{maxauthors=30, etalmode=truncate}

   \makeatletter
\newcommand*{\addFileDependency}[1]{
  \typeout{(#1)}
  \@addtofilelist{#1}
  \IfFileExists{#1}{}{\typeout{No file #1.}}
}
\makeatother

\newcommand*{\myexternaldocument}[1]{%
    \externaldocument{#1}%
    \addFileDependency{#1.tex}%
    \addFileDependency{#1.aux}%
} 

\myexternaldocument{supp}

\usepackage[version=3]{mhchem}

\newcommand{\etal}{\textit{et al.}}

\newenvironment{competing interests}
{
  \par\vspace{\baselineskip} \noindent
  \begin{Large}\textbf{Competing Interests}\end{Large} 
  \par \noindent\ignorespaces
}

\newenvironment{data availability}
{
  \par\vspace{\baselineskip}\noindent
  \begin{Large}\textbf{Data Availability}\end{Large}
  \par \noindent\ignorespaces
}

\newenvironment{author contribution}
{
  \par\vspace{\baselineskip}\noindent
  \begin{Large}{\textbf{Author Contribution}} \end{Large}
  \par \noindent\ignorespaces
}

\author{Shahid Sattar}
 \email{shahid.sattar@lnu.se}
\affiliation{Department of Mathematics and Physics,\\ Linnaeus University, SE-39231 Kalmar, Sweden}
\author{Roman Stepanov}
\affiliation{Department of Mathematics and Physics,\\ Linnaeus University, SE-39231 Kalmar, Sweden}
 \author{C. M. Canali}
\affiliation{Department of Mathematics and Physics,\\ Linnaeus University, SE-39231 Kalmar, Sweden}

\title{Topological Magneto-Optical Switching in Even-Layered MnBi$_2$Te$_4$}


\begin{document}

\begin{tocentry}
\includegraphics{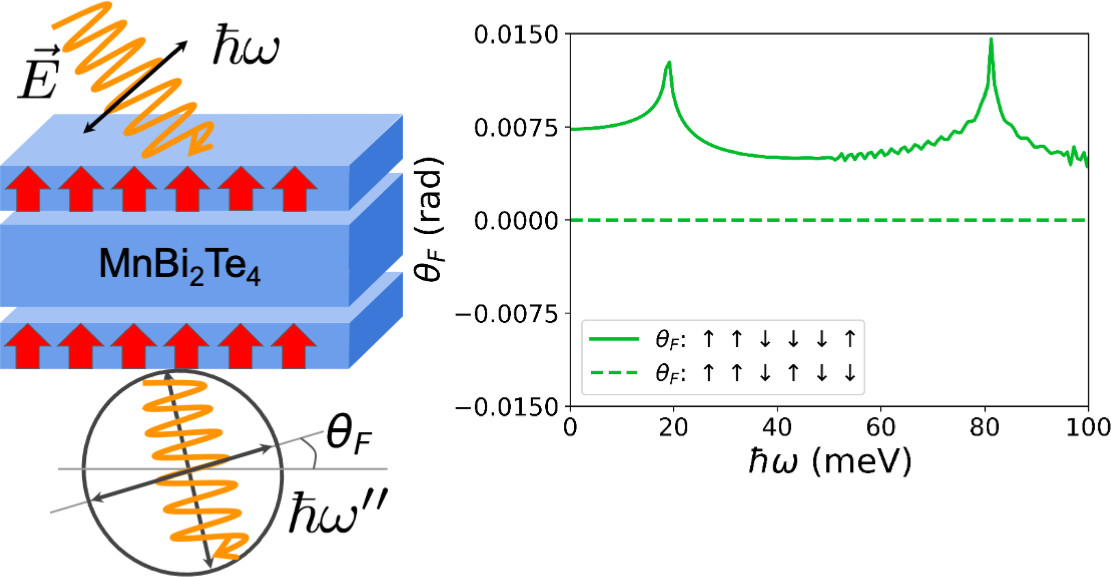}
\end{tocentry}

\begin{abstract} 

MnBi$_2$Te$_4$ (MBT) thin films provide a unique material platform in which magnetism, topology, and magneto-optical (MO) response can be tuned through layer-thickness and relative spin alignments. In this work, using a low-energy coupled Dirac cone model together with Wannier-based tight-binding Hamiltonian derived from \textit{ab-initio} calculations, we investigate topological MO switching in even-layered MBT films. We argue that the relative spin alignment of the outermost septuple-layers (SL) mainly controls the total Chern number, optical conductibility, and consequently, the MO response. For a 6-SL MBT thin film, we found that reversing the outermost-SL alignments from antiparallel to parallel switches the system from axion insulating state with $C=0$ and vanishing Faraday rotation to a Chern insulating state with $C=1$ and a quantized MO response, irrespective of $PT$-symmetry and net magnetization. Increasing thickness reveals an additional regime: while 8-SL MBT hosts only $C=0$ and $1$ states, a 12-SL MBT film supports a higher Chern number phase with $C=2$ with a doubled low-frequency Faraday rotation. Our results provide a thickness-dependent route to multilevel MO switching and establish MO spectroscopy as a direct probe of surface magnetism and topological order in MBT thin films. 

\end{abstract}

\section{Introduction}

Magnetism and topology in intrinsic magnetic topological insulator MnBi$_2$Te$_4$ (MBT) give rise to rich hierarchy of topological phases, most prominently Chern and axion insulating states \cite{zhang2019topological,li2019intrinsic,otrokov2019unique,li2024progress,vyazovskaya2025intrinsic}. In MBT thin films, it is widely recognized that these phases are particularly sensitive to layer-thickness \cite{zhao2021even,ovchinnikov2021intertwined,chen2024even,li2024fabrication}. For example, the quantum anomalous Hall effect is typically observed in odd-layer samples and is associated with a finite Chern number (${C}=1$), whereas even-layer films generally host a ${C}=0$ state linked to axion electrodynamics \cite{deng2020quantum,liu2020robust,liu2021magnetic,lin2022direct}. Recent experiments, howoever, have uncovered a cascade of field-driven spin-flip transitions \cite{lian2025antiferromagnetic}, highlighting that the microscopic spin arrangement within the constituent septuple layers (SLs) constitutes an additional and equally important control parameter, as suggested previously by theoretical works \cite{lei2021metamagnetism,lei2022quantum}. Since the low-energy electronic structure and topology is mainly governed by the exchange-driven gapped topological surface states \cite{lei2023kerr,sun2020analytical}, spin alignments together with film thickness provide a natural route for switching between distinct topological states \cite{liu2020robust,lei2021metamagnetism}. 

Unlike conventional probes, MO spectroscopy can directly access the topological electrodynamics of MBT thin films because the Faraday and Kerr responses are set by the frequency-dependent anomalous Hall conductivity of exchange-gapped Dirac surface states \cite{ghosh2024probing,lei2023kerr}. Several experiment works on MBT thin films have demonstrated the potential of optical probes for accessing axion-related electrodynamics, MO effects, and ultrafast spin dynamics \cite{qiu2025observation,bartram2023real}. Quantized magneto-terahertz effects (Faraday and Kerr rotations) reported in 6-SL and 7-SL MBT thin films \cite{han2025quantized} and helicity-dependent optical control of AFM order in a 6-SL MBT films \cite{qiu2023axion} further underscore the potential of MO probes for accessing and manipulating coupled magnetic and topological degrees of freedom in MBT thin films.

These developments therefore motivate the use of Faraday rotation as a accessible probe for disentangling how SL spin alignments, layer-resolved Chern contributions, and film thickness jointly determine the topological character of MBT thin films. Therefore, in this work, we focus on evaluating the MO response in the form of Faraday rotation angle, governed by the frequency-dependent anomalous Hall conductivity and providing details of underlying topology rather than merely serving as a probe of magnetization \cite{ghosh2024probing}. Our study involves employing a low-energy coupled Dirac cone model together with tight-binding models derived from first-principles calculations. We discuss possibility of achieving topological MO switching in even-layer MBT thin films driven by SL spin alignments and layer thickness. We argue that irrespective of the presence or absence of $PT$ symmetry and net magnetization, the decisive factor is the relative spin alignment of the outermost SLs. When the top and bottom SLs are aligned parallel, MBT thin films exhibit a quantized Faraday rotation associated with a total Chern number. In contrast, opposite surface spin alignment leads to a vanishing Faraday angle and an axion insulating phase with ${C}=0$. We further show that this pattern naturally extends to thicker even-layer samples, where higher Chern number states with ${C}=2$ emerge owing to additive BC contributions. Our results thus identify surface spin alignment as the microscopic mechanism governing topological MO switching in even-layered MBT and establish MO spectroscopy as a probe of surface magnetism and topological order in MBT thin films.

\section{Computational method}

We employed two complementary theoretical approaches in this study. First, we used a simplified coupled Dirac-cone model developed for MnBi$_2$Te$_4$ in Ref.\cite{lei2020magnetized,lei2023kerr}, which provides a minimal low-energy description of electronic structure, topology, magnetism, and finite-thickness effects in this material; the model details are presented in the next section. Second, we performed first-principles calculations within density functional theory (DFT) using the projector augmented wave method\,\cite{paw1,paw2} as implemented in the Vienna Ab-initio Simulation Package (VASP) \cite{vasp}. The exchange-correlation functional was treated within the generalized gradient approximation in the Perdew-Burke-Ernzerhof parametrization \cite{perdew1996generalized}. Spin-polarized calculations including spin-orbit coupling were performed with a plane-wave cutoff energy of 430 eV and a $\Gamma$-centered $12\times12\times1$ Monkhorst-Pack $k$-mesh for the self-consistent calculations. We solved Kohn-Sham equations iteratively until the total energy and atomic forces were converged to $10^{-7}$ eV and $10^{-3}$ eV/\AA, respectively. The on-site Coulomb interaction on Mn $d$ orbitals was treated within the Dudarev’s scheme \cite{dudarev1998electron} using $U_{eff} = 3.9$ eV. Long-range van der Waals (vdW) interactions were included using Grimme’s DFT-D3 dispersion correction \cite{grimme2010consistent}, and a vacuum slab of 15~\AA\ was introduced along the out-of-plane direction to eliminate spurious interactions.

The purpose of doing DFT calculations is to obtain a real-space tight binding (TB) model by parameterizing the self-consistent Kohn-Sham Hamiltonian using the Wannier90 package \cite{marzari1997maximally,mostofi2014updated,pizzi2020wannier90}. We used the VASP2WANNIER90 interface and included Mn$-d$, Bi$-p$ and Te$-p$ orbitals in generating the Wannier functions. Topological characteristics were computed using the WannierTools \cite{wu2018wanniertools} and WannierBerri packages \cite{tsirkin2021high}. For the computation of the optical conductivity, using both TB and analytical models, we used the Kubo-Greenwood formula \cite{kubo1957statistical,greenwood1958boltzmann}:
\begin{equation} 
\begin{aligned}
\sigma_{\alpha\beta}(\hbar\omega)=\frac{i e^2}{\hbar}
\int\!\frac{d^2k}{(2\pi)^2}\sum_{n m}
\frac{f_{n\mathbf k}-f_{m\mathbf k}}{E_{n\mathbf k}-E_{m\mathbf k}}
\times\frac{\big|\langle m\mathbf k|\partial_{k_x} H_{\mathbf k}|n\mathbf k\rangle\big|\big|\langle m\mathbf k|\partial_{k_y} H_{\mathbf k}|n\mathbf k\rangle\big|}
{E_{n\mathbf k}-E_{m\mathbf k}-(\hbar\omega+i\eta)},
\end{aligned} \label{eq-1}
\end{equation}
where $f_{n\mathbf k}$ is Fermi-Dirac distribution function giving band occupation probability, $\omega$ is the optical frequency, $\hbar$ is the reduced Planck’s constant, $n,m$ are band indices, $|n\mathbf k\rangle$ and $|m\mathbf k\rangle$ are Bloch states and  $E_{n\mathbf k}$ and $E_{m\mathbf k}$ are the corresponding band energies, and $\eta$ is a disorder broadening parameter. The Faraday rotation was then obtained from the optical conductivity tensor, as described in the main text.

\section{Results and Discussion}

Bulk MnBi$_2$Te$_4$ (MBT) preserves the combined $PT$-symmetry, where $P$ denotes spatial inversion (exchanging the magnetic sublattices in adjacent SLs) and $T$ is time-reversal invariance. For even-layered MBT thin films, however, the survival of this symmetry is no longer guaranteed by crystal structure alone, but depends sensitively on the relative spin orientation of the neighboring SLs. The enumeration of all possible spin-configurations for a 6-SL film (table S1), including both compensated and uncompensated cases, shows that only four configurations preserve $PT$-symmetry, all of which belong to the fully compensated class. This distinction is central to the emergence of axion and Chern insulating phases, which both require a sizable exchange gap but differ fundamentally in their side-wall transport, i.e., whether the edge-states remain gapped or become conducting. Accordingly, the spin configuration in a thin-film directly controls the behavior of Dirac surface states, with their gaps and hybridization determined by interlayer/intralayer hopping and magnetic exchange. These considerations motivate the use of the minimal coupled Dirac-cone model, introduced by Chao \etal \cite{lei2020magnetized,lei2023kerr}, as a first choice since it effectively captures the interdependence of magnetism, topology, and finite-size effects in MBT thin films.

\begin{figure*}[!htb]
    \centering
    \includegraphics[width=14cm]{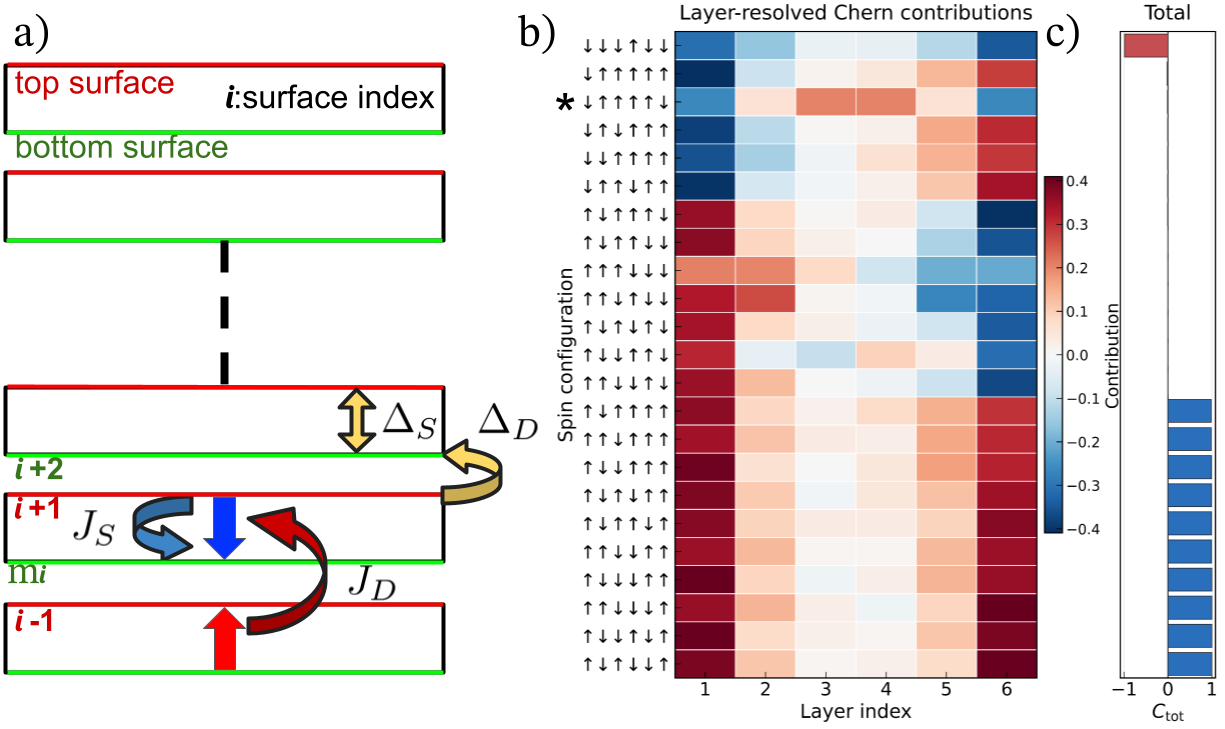}
    \caption{(a) Schematic illustration of the coupled Dirac cones model for even-layered MBT thin films. The top and bottom surface of each septuple layer (SL) is shown in red and green colors, respectively and correspond to opposite signs of the Dirac kinetic energy term. The arrows denote the four model parameters: $J_D$ is the magnetic exchange from the neighboring SL, $J_S$ is the intralayer exchange, and $\Delta_S$ ($\Delta_D$) is the intralayer/interlayer hybridization. The red and bright blue arrows indicate up and down spin alignments. (b) Heat map of the layer-resolved Chern contributions for different spin alignments in a 6-SL MBT film. (c) Sum of all contributions (i.e., total Chern number $C_{\mathrm{tot}}$) for the respective spin configurations.}
    \label{fig:fig1}
\end{figure*}

Fig. \ref{fig:fig1}(a) presents a schematic representation of the model for an even-layered MBT thin-film of arbitrary thickness. In this model, each SL hosts two Dirac cones, which are spatially localized at the top and bottom surface, denoted by the surface index $i$. The Dirac cones within the same SL experience an intralayer hybridization $\Delta_S$, while those in adjacent SLs are coupled via an interlayer hybridization $\Delta_D$. Similarly, the magnetic exchange coupling between a given Dirac surface state and the local Mn magnetic moments is parameterized by $J_S$ for interactions within the same SL and by $J_D$ for interactions with neighboring SLs. Thus, for a film consisting of $N$ SLs, the Hamiltonian is constructed in a basis of $2N$ Dirac cones, explicitly incorporating the spin degree of freedom and is given by the following expression:
\begin{align}
&H = \sum_{\mathbf{k}_\perp,i,j}
\Big[
\big((-1)^{i}\,\hbar v_D \,(\hat{\mathbf z}\!\times\!\boldsymbol{\sigma})\!\cdot\!\mathbf{k}_\perp
+ m_i\,\sigma_z \big)\,\delta_{ij}
\;+\Delta_{ij}\,(1-\delta_{ij})
\Big]\,c^\dagger_{\mathbf{k}_\perp i}\,c_{\mathbf{k}_\perp j}
\end{align}
Here, $\mathbf{k}_\perp$ is the in-plane two-dimensional crystal momentum, whereas $c^\dagger_{\mathbf{k}_\perp i}$ and $c_{\mathbf{k}_\perp j}$ are the electron creation and annihilation operators in the $i$-th and $j$-th Dirac cones, respectively. The indices $i$ and $j$ are assigned sequentially across SLs, with odd indices corresponding to the upper surface of a given SL and even indices to the lower surface. The exchange and hybridization parameters employed in this model were determined by fitting the exchange gap obtained from DFT calculations of a 6-SL MBT thin film ($E_{\text{g}}= 59$ meV) to the corresponding gap of the present model. This procedure yields magnetic exchange parameters $J_S=40$ meV (intralayer) and $J_D= 46$ meV (interlayer). Hybridization is incorporated via two matrix element types: intralayer coupling $\Delta_S=80$ meV, which connects the upper and lower Dirac cones within the same SL, and interlayer coupling $\Delta_D=-153$ meV, which connects the lower cone of the $N$-th SL to the upper cone of the $(N+1)$-th SL.

A 6-SL MBT thin film ($N=6$) exhibits net-zero magnetization in its ground state but possesses $C(N,k)/2=10$ unique compensated spin configurations with $k=3$ specifies the number of spin-up (or spin-down) Mn moments. Moreover, under an applied external magnetic field, numerous uncompensated configurations having finite non-zero magnetization can be achieved (see Table S1 for a full list). Consequently, we investigated a broad manifold of both compensated and uncompensated spin alignments, a subset of which is illustrated in Fig. \ref{fig:fig1}(b). We calculated layer-resolved Chern contributions to each SL by weighting the Berry curvature (BC) of each occupied state by its projection onto the corresponding SL, according to the following definition:

\begin{equation}
C_l =
\frac{1}{2\pi}
\sum_{n \in \mathrm{occ}}
\int d^2k \,
W_{n,l}(\mathbf{k})\,\Omega_n(\mathbf{k}),
\end{equation} 

where $W_{n,l}(\mathbf{k})$ is the weight of the n-th eigenstate on the l-th septuple layer and $\Omega_n(\mathbf{k})$ is the Berry curvature. In addition, refer to Fig. \ref{fig:fig1}(c), the total Chern number ($C$) is computed by integrating the BC over the 2D Brillouin zone for each case.

For the fully compensated configurations that preserve $PT$-symmetry, layer-resolved Chern contributions exhibit an antisymmetric distribution about the film center, leading to an exact cancellation of BC and a vanishing total Chern number ($C_{tot}=0$). Interestingly, for the remaining compensated and uncompensated spin-configurations, irrespective of the presence of $PT$-symmetry or net magnetization, an opposite spin alignment at the top and bottom SL can still yield the same ($C_{tot}=0$) as depicted in Fig. \ref{fig:fig1}(b)-c) for several cases. In contrast, parallel spin-alignment at the outermost SLs yields a finite Chern number ($C_{tot}=1$), owing to the addition of Dirac mass terms of the same sign on the two surfaces (see layer-resolved contributions). We notice that the dominant Chern contributions are localized near the outermost SLs, consistent with the surface origin of the Dirac states, while their gradual variation across the thin-film highlights the role of interlayer hybridization in redistributing BC. Hence, these results establish a direct correspondence between spin alignment of the outermost SLs and the resulting topological response; i.e, anti-parallel alignment realizes an axion-insulating state ($C_{tot}=0$), whereas parallel alignment drives the system into a Chern insulating phase ($C_{tot}=1$). In order to confirm this, we utilized real-space TB models obtained from DFT calculations for the two representative cases and plot quasi-1D side-wall edge states (see Supplementary Fig. S1). The existence of gapped and chiral side-wall edge states for axion and Chern insulator further support our analysis. An exception, however, arises for the spin-configuration $\downarrow\uparrow\uparrow\uparrow\uparrow\downarrow$ (third row in Fig. \ref{fig:fig1}(b)), which corresponds to a topological phase transition point. In this case, the contributions from the outermost SLs are no longer dominant (see Supplementary Fig. S2). For the special spin configuration at the changeover point ($C=0 \leftrightarrow C=1$)), the exchange gap vanishes ($E_{\text{g}}= 0$ meV) and hence the Chern number becomes ill-defined. 

\begin{figure*}[!]
    \centering
    \includegraphics[width=16cm]{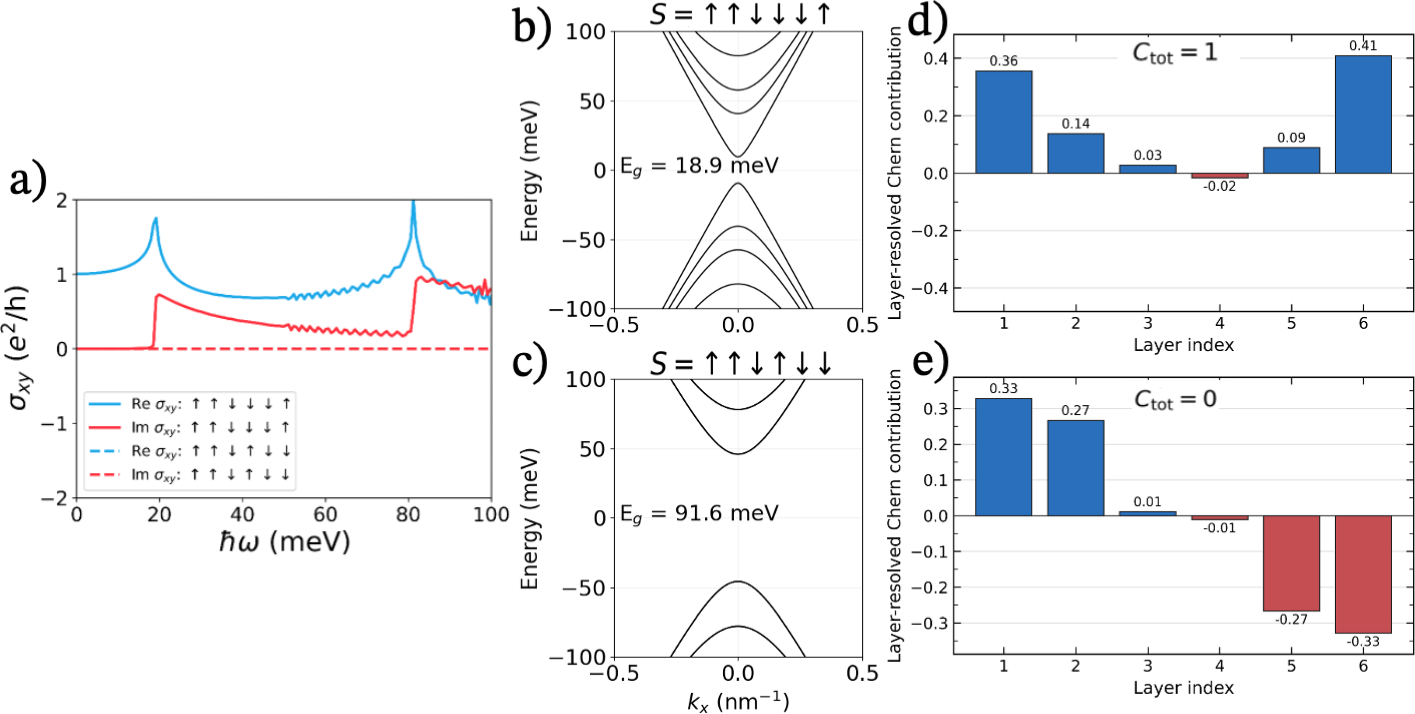}
        \caption{Frequency dependent anomalous Hall conductivity ($\sigma_{xy}$) for representative 6-SL spin configurations. (b-c) Corresponding low-energy band structures in the vicinity of the $\Gamma$ point with band gaps E$_g$=18.9 meV and E$_g$=91.6 meV for C$_{tot}$=1 and C$_{tot}$=0, respectively. (d-e) Layer-by-layer contributions to the Chern number for the same configurations.}
    \label{fig:fig2}
\end{figure*}

Fig. \ref{fig:fig2} presents the frequency dependent anomalous Hall conductivity ($\sigma_{xy}$), electronic band structure with the magnitude of exchange gap ($E_{g}$), and layer-resolved Chern contribution for the two representative spin configurations of a 6-SL MBT thin film, providing a unified picture of the distinct topological regimes. For the configuration with antiparallel spin alignment at the outermost SLs, the system remains insulating with a sizable exchange gap. We observe both real and imaginary parts of optical conductivity to exhibit a vanishing low-energy response (shown as dotted red and blue lines in Fig. \ref{fig:fig2}(a)), consistent with an axion insulating phase characterized by $C_{tot}=0$. In contrast, for the parallel SLs alignments on the surface, we found a quantized optical response (Re $\sigma_{xy}=1$: solid red line, Im $\sigma_{xy}=0$: solid blue line) reflecting the onset of a Chern insulating state having $C_{tot}=1$. Refer to Fig. \ref{fig:fig2}(b)-(c), the reduction (increase) of exchange gap to $18.9$ ($91.6$) meV compared to AFM ground state ($59$ meV) is due to the spin-accumulation in the center (surface) SLs which is consistent with recent experimental findings \cite{lian2025antiferromagnetic}. The respective SL-resolved Chern contribution given in Fig. \ref{fig:fig2}(d)-(e) corroborate these findings, i.e., top and bottom SL dominance and showing same-sign surface contributions addition in the latter case and cancellation in the former. These results collectively demonstrate that the optical and topological responses are governed by the spin alignment of the outermost SLs in an even-layered MBT thin film, consistent with the coupled Dirac-cone description.

\begin{figure*}[!]
    \centering
    \includegraphics[width=14cm]{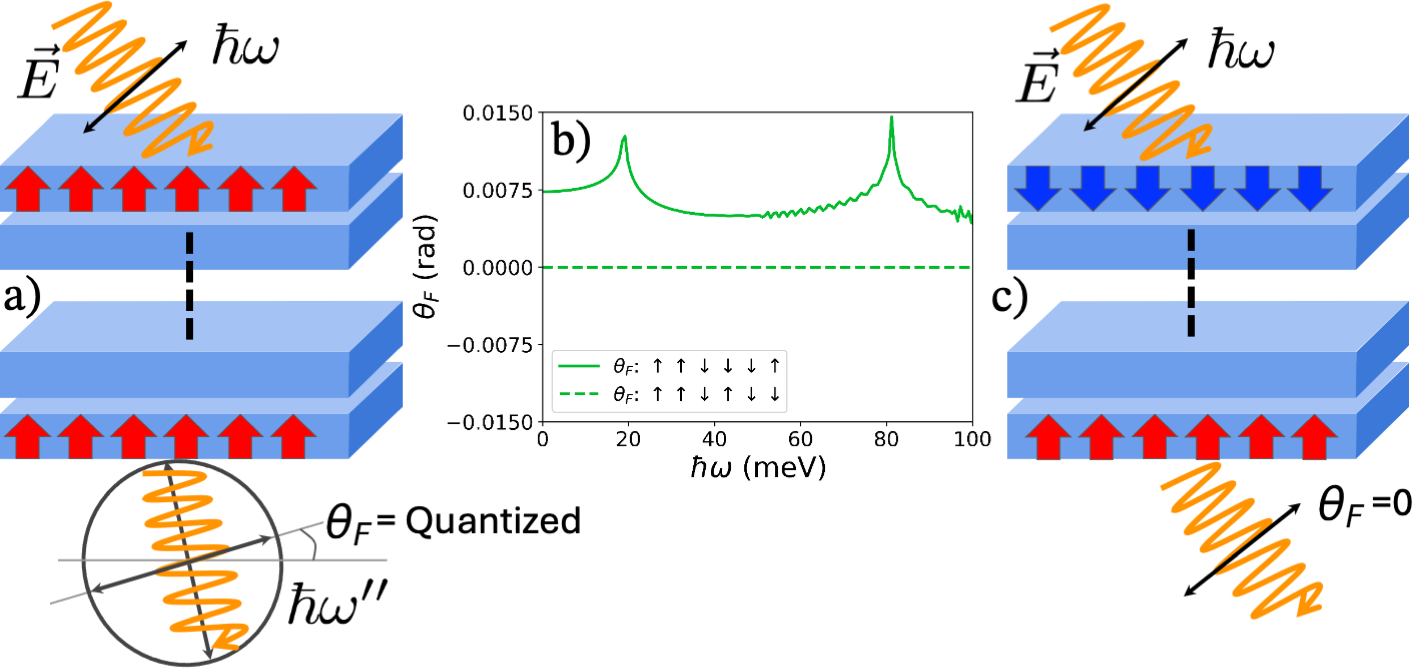}
        \caption{Topological MO switching in a 6-SL MnBi$_2$Te$_4$ thin film. (a) Schematic of incident light ($\hbar \omega$) falling on a spin configuration with parallel magnetization at the outermost SLs, resulting in light transmission ($\hbar \omega ''$) and a quantized Faraday rotation angle ($\theta_F$). (b) Frequency-dependent Faraday angle for two representative spin configurations: the solid line corresponds to parallel SL alignment ($C=1$), exhibiting quantized MO response, while the dashed line corresponds to antiparallel alignment ($C=0$) showing a vanishing response. (c) Illustration of the antiparallel configuration, where opposite magnetization at the outermost SLs leads to zero Faraday rotation angle ($\theta_F = 0$).}
    \label{fig:fig3}
\end{figure*}

Fig. \ref{fig:fig3} illustrates the central mechanism of topological MO switching in even-layered MBT, driven solely by the spin alignment of the outermost SLs. We consider a representative case of parallel spin alignments on the outermost SLs as shown in Fig. \ref{fig:fig3}(a). When linearly polarized light of energy $\hbar\omega$ is incident on this system, the finite anomalous Hall conductivity ($\sigma_{xy}$) induces different phase shifts for the left- and right-circularly polarized components, resulting in a rotation of the polarization plane upon transmission. For a MBT thin film of finite thickness $d$ ($d\ll\lambda$, with $\lambda$ being the wavelength of incident light) placed between two dielectrics having refractive indices $n_t=\sqrt{\varepsilon_1}$ (top) and $n_b=\sqrt{\varepsilon_2}$ (bottom), the MO response is characterized by 2D conductive layers with surface conductivity $\sigma$ (measured in units of $e^2/h$). Imposing the boundary conditions allows us to determine the transmission ($E_{x/y}^{t}$) and reflection field ($E_{x/y}^{r}$). We then express these fields in a linear polarization basis:
\begin{align}
\binom{E_x^{t}}{E_y^{t}}
&= \frac{1}{M}
\binom{\,2 n_t\!\left(n_t+n_b+2\alpha\,\sigma_{xx}\right)\,}{-\,4\alpha n_t\,\sigma_{xy}},\label{eq-2}\\[4pt]
\binom{E_x^{r}}{E_y^{r}}
&= \frac{1}{M}
\binom{\,n_t^{2}-\!\left(n_b+2\alpha\,\sigma_{xx}\right)^{2}-\left(2\alpha\,\sigma_{xy}\right)^{2}\,}{-\,4\alpha n_t\,\sigma_{xy}},\label{eq-3}
\end{align}
where,
$M=\left(n_t+n_b+2\alpha\,\sigma_{xx}\right)^{2}+\left(2\alpha\,\sigma_{xy}\right)^{2}$. Using Eq.~\eqref{eq-2} and \eqref{eq-3} together with the optical conductivity tensor computed via the Kubo-Greenwood formula of Eq.~\eqref{eq-1}, we finally obtain expressions for the angles $\theta_F$ as follows:
\begin{equation}
\theta_F=\tfrac{1}{2}\!\left(\arg E_{+}^{t}-\arg E_{-}^{t}\right),\label{eq-4}
\end{equation}
where $E_{\pm}^{\,t,r}=E_x^{\,t,r}\pm i\,E_y^{\,t,r}$. In the low-frequency limit ($\hbar \omega << \textbf{$E_{\rm gap}$}$), the Faraday rotation angle ($\theta_F$) acquires the ``quantized" value of $\theta_F=C \;$tan$^{-1}\alpha$ ($\alpha$ is the fine structure constant) reflecting the underlying topological character of the system. Within the coupled Dirac-cone model, parallel spin alignment of outmost SLs produces Dirac masses of the same sign, leading to the addition of BC and a finite Chern number (${C}=1$). This behavior is captured in Fig. \ref{fig:fig3}(b), where the frequency-dependent Faraday rotation angle exhibits a finite response for the parallel configuration (solid line), showing a quantized plateau at low frequencies. We notice that MO response is dictated by the topology of the electronic structure rather than microscopic details of MBT thin film. In contrast, when the outermost SLs are antiparallel, as illustrated in Fig. \ref{fig:fig3}(c), the Dirac masses acquire opposite signs, resulting in cancellation of surface contributions and a vanishing Chern number (${C}= 0$). Consequently, the Faraday rotation is absent across the entire frequency range, as shown by the dashed line in Fig. \ref{fig:fig3}(b). Together, these results demonstrate that switching the relative spin alignments of the outermost SLs enables a direct transition between the quantized and a vanishing MO response, thereby realizing topological MO switching in even-layered MBT.

\begin{figure*}[!]
    \centering
    \includegraphics[width=16cm]{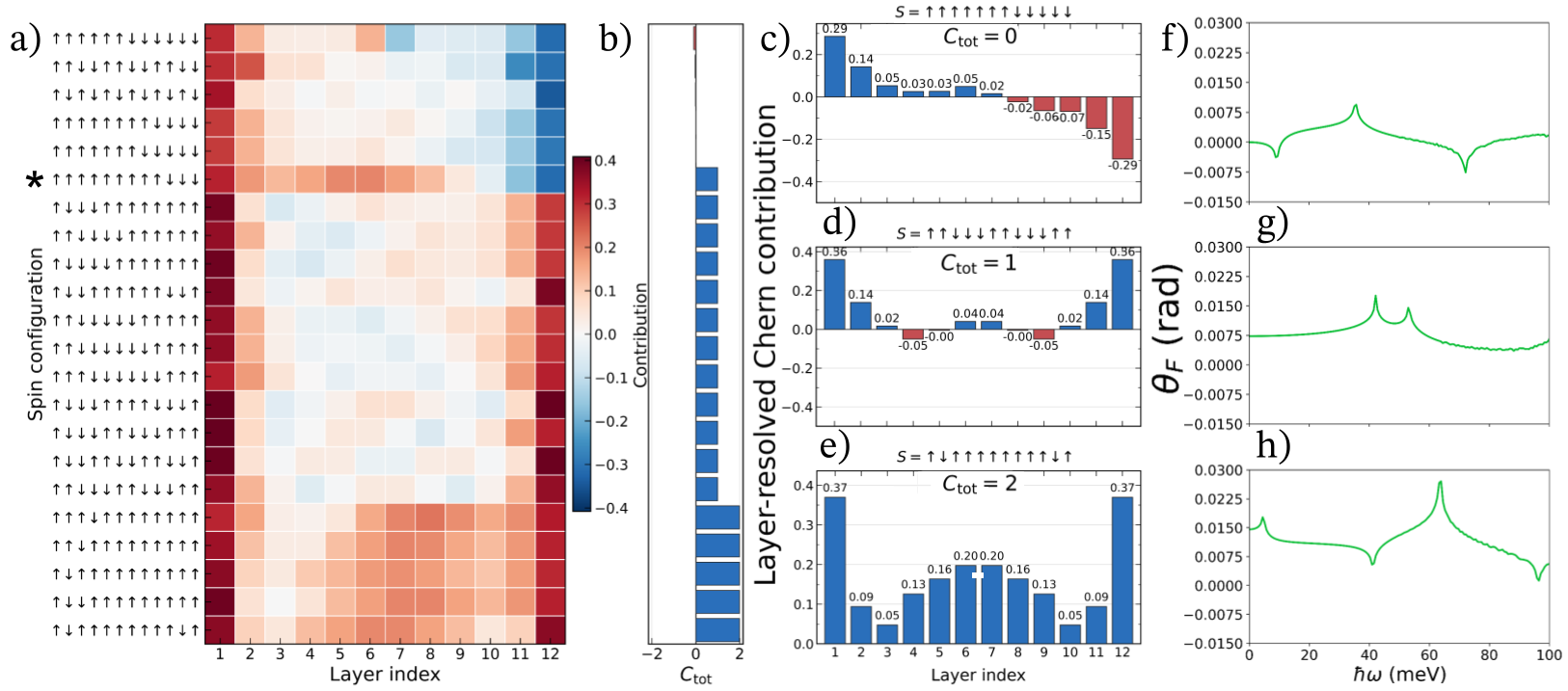}
    \caption{(a) Heat map of layer-resolved Chern contributions across different spin configurations considered for a 12-SL MBT thin film. (b) Corresponding total Chern number ($C_{tot}$) map, identifying $C_{tot}=0, 1,\,\text{and}$\,2 phases. (c-e) Representative layer-resolved Chern contributions for $C_{tot}=0, 1,\,\text{and}$\,2 states, respectively. (f-h) Frequency-dependence of the Faraday rotation angle ($\theta_F$) for the same three configurations. The $C_{tot}=2$ phase also exhibits quantized MO response with doubled low-frequency $\theta_F$, consistent with higher Chern number value, relative to $C_{tot}=1$ phase.}
    \label{fig:fig4}
\end{figure*}

We now turn our attention to thicker MBT samples which promise to host higher-Chern number (HCN) states (${C}=2$ and beyond) in addition to the possibility of having ${C}= 1,$ and $0$ Chern and axion insulating phases, respectively. The existence of HCN states is well motivated by several theoretical works on magnetic TIs \cite{wang2013quantum, fang2014large}, engineered MBT multilayers \cite{bosnar2023high} and mixed-stacking films \cite{li2025high}. Experimentally, Ge\,\etal\cite{ge2020high} reported a HCN state having ${C}=2$ Hall plateau in $9$- and $10$-SL MBT devices under applied magnetic field, establishing that the signature of HCN can emerge in this material platform. Moreover, thicker MBT thin films could potentially be used for tuning the Chern number as demonstrated earlier for magnetically-doped TI \cite{zhao2020tuning}. In this context, looking at the thickness dependence of total and projected Chern contribution, we first investigate different spin configurations in 8-SL MBT thin film which are given in Supplementary Fig. S3. Notably, our model calculations yield only ${C}= 0$ and $1$ states with no evidence of a $C=2$ state, indicating that the HCN states could emerge only beyond a critical film thickness. Consequently, the 12-SL case provides a natural platform for realizing HCN states and allows us to predict MO response as a probe in experiments. 

Figure \ref{fig:fig4}(a) presents a set of spin configurations for a 12-SL MBT thin film, together with the corresponding layer-resolved Chern contributions, extending the analysis performed earlier for the 6-SL and 8-SL cases to a thicker film with additional topological channels and the emergence of HCN states. In addition, the total Chern number map ($C_{tot}$) of Fig. \ref{fig:fig4}(b) separate multiple spin configurations into $C_{tot}=0,1,$ and $2$ sectors. This should be contrasted with earlier thicknesses, where an analogous analysis yields no HCN $C_{tot}=2$ phase, highlighting that the higher Chern response is not a generic feature of even-layered MBT thin films. Refer to Fig. \ref{fig:fig4}(c), positive and negative layer-resolved Chern contributions compensate across the slab and especially at the outermost SLs, resulting in vanishing topological response with $C_{tot}=0$ despite finite BC weights. In the $C_{tot}=1$ phase shown in Fig. \ref{fig:fig4}(d), this cancellation is incomplete because the outermost SLs and near-surfaces contribute with an overall common sign, leaving a net unit Chern number. Most importantly, the HCN $C_{tot}=2$ state in Fig. \ref{fig:fig4}(e) exhibits commutative addition of Chern contribution havingthe largest weight concentrated at the two outermost SLs. Moreover, finite positive contribution from multiple SLs extending into adjacent inner SLs effectively accommodate more than one Dirac channel with the same mass sign. These results are consistent with an earlier theoretical proposal that, for a given magnetic moment strength, a larger film thickness leads to a higher Chern number ($C=2$ in this case) \cite{wang2013quantum}. The MO response given in Fig. \ref{fig:fig4}(f-h) mirrors this hierarchy at low frequencies, i.e., the Faraday rotation angle ($\theta_F$) is quenched for $C_{tot}=0$, becomes quantized to $\theta_F=C \;$tan$^{-1}\alpha$ for $C_{tot}=1$ phase and is doubled for HCN $C_{tot}=2$ state. Taken together, these results show that 12-SL MBT supports not only topological MO switching between axion- and Chern-insulating phases, but also a higher MO state distinguished by a double Faraday rotation signature.

\section{Conclusion}

In summary, we have shown that even-layered MBT thin films of varying thickness provide a unique avenue for topological MO switching driven entirely by the spin alignment of the outermost SLs. Combining a coupled Dirac-cone model and \textit{ab-initio} calculations, we have demonstrated that the relative spin configuration of constituent SLs controls the sign and additivity of the surface Dirac mass terms, thereby determining the layer-resolved and total Chern contributions, anomalous Hall conductivity and MO response. For 6-SL MBT, reversing the spin alignment of the outermost SLs from antiparallel to parallel switches the system from an axion insulating state with $C_{tot}=0$ and vanishing Faraday rotation to a Chern insulating state with $C_{\rm tot}=1$ and a quantized MO response. Extending this analysis to thicker films reveals a pronounced thickness dependence: while 8-SL MBT supports only $C_{\rm tot}=0$ and $1$ states, a 12-SL film can host a HCN state with $C_{\rm tot}=2$. This phase arises from constructive BC contributions from multiple gapped Dirac channels and manifests as a low-frequency Faraday angle twice that of $C_{\rm tot}=1$ state, consistent with the expected Chern-number dependent MO response. Taken together, these results establish even-layered MBT as a unique topological system with spin alignment and thickness-dependent Chern number and associated MO response. Our results provides a microscopic mechanism for reversible switching between axion, Chern, and HCN states, highlighting MBT as a promising platform for multilevel topological MO response. 

\begin{acknowledgement}
Fruitful discussions with A. H. MacDonald and Chao Lei are greatly acknowledged. We thank Swedish Research Council (grant no: VR 2021-04622) for financial support. The computations were enabled by resources provided by the National Academic Infrastructure for Supercomputing in Sweden (NAISS) partially funded by the Swedish Research Council through grant agreement no. 2022-06725 and the Centre for Scientific and Technical Computing at Lund University (LUNARC).
\end{acknowledgement}

\begin{competing interests}
The Authors declare no competing financial or non-financial interests.
\end{competing interests}

\begin{data availability}
The data that support the findings of this study are available from the corresponding
author upon reasonable request.
\end{data availability}

\begin{author contribution}
S. Sattar and R. Stepanov performed the calculations, C. M. Canali analyzed the results.  All authors reviewed the manuscript. 
\end{author contribution}

\bibliography{main.bib}

\begin{suppinfo}
The Supplementary Information contains (1) spin alignments in a 6-SL MBT thin film, (2) Fig. S2 containing frequency-dependent anomalous Hall conductivity and Chern contributions for the exceptional spin configuration, (3) Fig. S3 showing heat map of layer-resolved Chern contributions across different spin configurations considered for a 8-SL MBT thin film. 
\end{suppinfo}

\end{document}